# Epitaxial strain induced magnetic transitions and phonon instabilities of the tetragonal SrRuO₃

He He,[1] Hang-Chen Ding,[1] Yong-Chao Gao,[1] Shi-Jing Gong,[1,2,*] Xiangang Wan,[3] Chun-Gang Duan[1,2]

[1] *Key Laboratory of Polar Materials and Devices, Ministry of Education, East China Normal University, Shanghai 200062, China*

[2] *National Laboratory for Infrared Physics, Chinese Academy of Sciences, Shanghai 200083, China*

[3] *National Laboratory of Solid State Microstructures and Departmentof Physics, NanjingUniversity, Nanjing 210093, China*

Using density-functional theory calculations, we investigate the magnetic as well as the dynamical properties of tetragonal SrRuO₃ (SRO) under the influence of epitaxial strain. It is found that both the tensile and compressive strain in the *xy*-plane could induce the abrupt change in the magnetic moment of Ru atom. In particular, under the in-plane ~4% compressive strain, a ferromagnetic to nonmagnetic transition is induced. Whereas for the tensile strain larger than 3%, the Ru magnetic moment drops gradually with the increase of the strain, exhibiting a weak ferromagnetic state. We find that such magnetic transitions could be qualitatively explained by the Stoner model. In addition, frozen phonon calculations at Γ point reveal structural instabilities could occur under both compressive and tensile strains. Such instabilities are very similar to those of the ferroelectric perovskite oxides, even though SRO remains to be metallic in the range we studied. These might have influence on the physical properties of oxide supercells taking SRO as constituent.

In recent decades, the transition-metal perovskite oxides have been intensively investigated due to their fascinating properties, such as superconducting[1], ferroelectric,[2] and multiferroic[3,4] properties. Among them, strontium ruthenate (SrRuO₃, SRO), as one of the few conductive perovskite oxides adopted in the increasingly sophisticated epitaxial thin film technology, has attracted even more attention.[5] Currently, the theoretical research interest of SRO mainly focuses on the effects of the octahedral rotation and tilting as well as the electron correlations on the electronic, magnetic and transport properties of SRO.[6-14] Several different mechanisms have been proposed to explain the metal-insulating transition and the magnetic transition, e.g. correlation effects, spin-orbit coupling and magnetostructural coupling.[9,15-18] Guiding by its unique properties, SRO has a good perspective of practical applications, including multiferroic devices,[3] field effect devices,[19] ferroelectric capacitors,[20,21] magnetic and multiferroic tunneling junctions,[22-24] and so on.

The bulk SRO, with orthorhombic structure, keeps ferromagnetism up to $T_C$ ~160 K[25], with the spontaneous magnetic moment $M_s$=1.1-1.3$\mu_B$/f.u arising from the Ru 4d electrons. Due to the increasing research interest on SRO as the constituent of perovskite heterostructures or superstructures,[22,23,26,27] it is more interesting to study the SRO in the tetragonal phase, which generally arises from the epitaxial strain from the substrate. In fact, it has been reported that the transitions of the electronic band and magnetic structure of SRO can be induced by various facts, such as the structure distortion, the electron-electron correlations in 4d Ru,[16] the growth temperature,[18] and the thickness of the film.[28-30] Therefore it is particularly important to understand the mechanism of the magnetic and electronic transition in the SRO ultrathin film.

In this paper, we investigate the evolution of magnetic properties of tetragonal SRO with the in-plane lattice constant *a* varying within the range 3.70 Å ~ 4.22 Å. The calculations reveal two critical lattice constants (*a*=3.85 Å and 4.11 Å), at which magnetic moment of Ru atom experiences an abrupt change. When 3.85 Å < *a* < 4.11 Å, the magnetic moment of Ru atom gradually increases with the increasing lattice constant *a*. Under the in-plane compressive strain with *a* ~ 3.85 Å, a ferromagnetic to nonmagnetic transition is induced. When *a* ~ 4.11 Å, magnetic moment of Ru atom sharply decreases to a small value. To distinguish these two ferromagnetic states, we call the one with larger magnetic moment strong ferromagnetic state, and the other weak ferromagnetic state. Detailed electronic structure analyses, together with the Stoner criterion are adopted to find the cause of these different magnetic states.

The calculations are performed within density-functional theory (DFT) using the projector augmented wave (PAW) method implemented in the Vienna Ab-Initio Simulation Package (VASP).[31-33] The exchange-correlation potential is treated in the generalized gradient approximation (GGA).[34] We use the energy cut-off of 500 eV for the plane wave expansion of the PAWs and a 10×10×10 Γ centered *k*-point grid in the self-consistent calculations. The Brillouin zone integrations are calculated using the tetrahedron method with Blöchl corrections.[35]

In all the following discussion, we take the theoretical lattice constant (*a* = 3.99 Å) for cubic SRO as a reference. According to our calculations, the magnetic moment of Ru in the cubic SRO is about 1.35$\mu_B$, agreeing well with previous reports.[7,10,14] Then we change the in-plane lattice constant *a* within the range 3.70 Å ~ 4.22 Å, which is equivalent to applying in-plane strain from about -7% to +6%. The lattice constant *c* is optimized for each *a*. Figure 1 shows the magnetic moment of Ru atom versus the in-plane lattice parameter $a$ of SRO. It is clear that there are two abrupt changes in the magnetic moment of Ru atom at two critical values of $a$: one is around 3.85 Å, and the other 4.11 Å. As the magnetic moment of Ru accounts for about 75% of the total magnetic moment of SRO, the solid squares in Fig. 1 also represents the trend of the magnetic moment of the whole system. When $a$ is larger than 4.11 Å or smaller than 3.85 Å, the magnetic moment of Ru atom experiences a sharp decrease. Especially for $a \sim 3.85$ Å, the compressive strain induces a ferromagnetic to nonmagnetic transition. When $a$ is within the range 3.85 Å ~ 4.11 Å, the magnetic moment of Ru atoms gradually increase with the increasing of $a$. Also in Fig. 1, we plot the ratio $c/a$ versus the lattice constant $a$ (see solid circles). It can be clearly seen that the $c/a$ curve is discontinuous, which also has anomalies at the same critical points as the curve of Ru's magnetic moment.

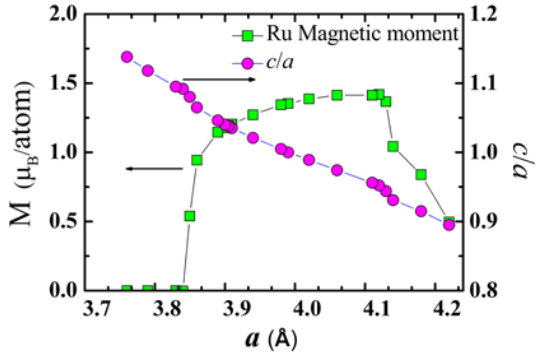

FIG.1. (color online) The magnetic moment of Ru atom (green solid squares) and the ratio $c/a$ (magenta solid circles) as a function of the in-plane lattice parameter $a$.

To well understand the magnetic transitions discussed above, we plot the distributions of the spin density in the $xz$ plane in Fig. 2, in which Ru atom is surrounded by four O atoms. Two different lattice constants ($a$=3.99 Å and 4.22 Å) are considered, which represent two different magnetic phases, i.e. strong and weak ferromagnetic states, respectively. We can clearly see from the shape of the spin density that it is $t_{2g}$ ($d_{xy,xz,yz}$) states that contribute most to the magnetic moment of the system. It is also obvious that the spin density in the cubic SRO ($a$=$c$=3.99 Å) is much stronger than that in the tetragonal SRO ($a$=4.22 Å, $c$=3.79 Å), and spin density distributions in the tetragonal SRO show obvious anisotropy along the $x$ and $z$ directions.

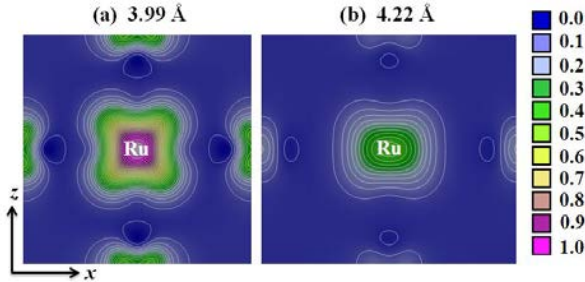

FIG.2. (color online) Spin density distribution in the $xz$ plane. The lattice parameter $a$ is equal to 3.99 Å and 4.22 Å, respectively.

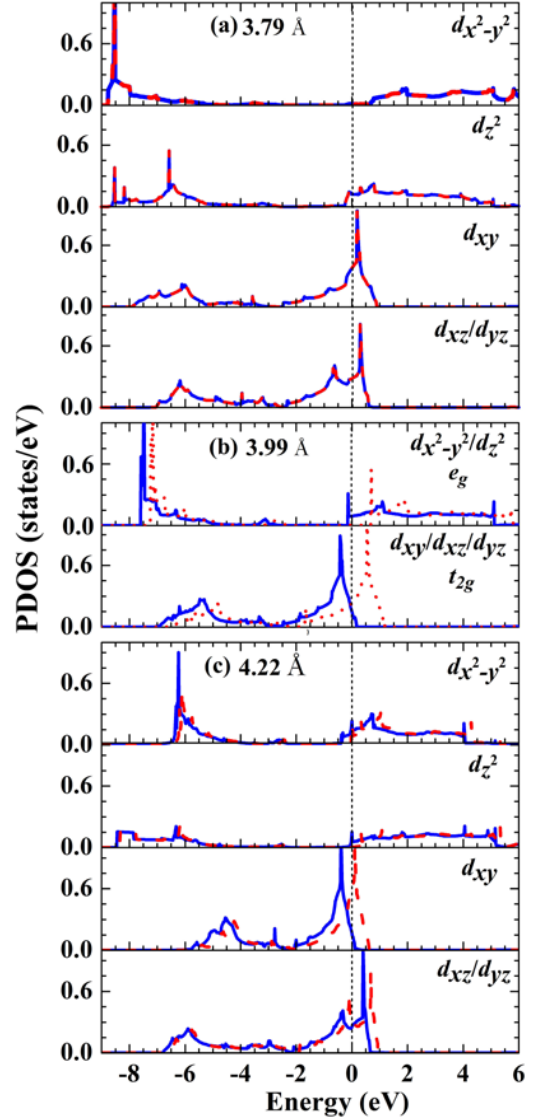

FIG.3. (color online) Spin and orbital-resolved partial density of states (PDOS) of Ru atom for SRO with the in-plane lattice parameter $a$ =3.79 Å (a), 3.99 Å (b), 4.22 Å (c). The spin up and spin down states are shown as blue straight lines and red dash dotted lines, respectively. Note that in (a), the spin up and spin down plots are identical. The black short dashed line marks the Fermi level.

The magnetic transition can be seen more clearly if we investigate spin and orbital-resolved partial density of states (PDOS) of Ru atom, which are shown in Figs. 3(a)-(c). Three in-plane lattice constants ($a$=3.79 Å, 3.99 Å, and 4.22 Å) are considered. For cubic SRO ($a$=3.99 Å), the three $t_{2g}$ orbitals ($d_{xz}$, $d_{yz}$, and $d_{xy}$) are degenerate, and the splitting between the up- and down-spin states is large, as can be clearly seen in Fig. 3(b). The double degenerate $e_g$ orbitals ($d_{x^2-y^2}$ and $d_{z^2}$) have higher energy than $t_{2g}$ orbitals, yet they have significant densities at around 7.5 eV below the Fermi level, due to the hybridization with $p$ orbitals of O atoms. This hybridization, however, contributes little to the magnetic moment of the system, as the occupation of $e_g$ states is almost identical for spin-up and spin-down channels. Therefore the magnetic moment of Ru atom in the cubic SRO is mainly from the $t_{2g}$ orbitals (about 95%).

For tetragonal SRO ($a$=3.79 Å and 4.22 Å), the $d_{xz}$ and $d_{yz}$ orbitals are still degenerate due to the tetragonal symmetry, yet the $d_{xy}$ orbital now has different energy. Under the compressive strain [$a$=3.79 Å, Fig. 3(a)], due to the shorter bond length, the hybridizations between Ru $d_{xy}$ states and in-plane O 2$p$ states enhance notably. Band-width of the $d_{xy}$ orbital becomes about 9.0 eV, much larger than that for $a$=3.99 Å (~7.1 eV). This generally indicates more effective electron hopping and weaker onsite electron-electron interactions. The band-width of the $d_{xz}/d_{yz}$ orbital is about 7.5 eV, which does not increase much due to the enlarged distance between Ru and atop O atoms. As for the tensile strain [$a$=4.22 Å, Fig. 3(c)], the situation is very different. Band-width of $d_{xy}$ states (~6 eV) becomes smaller and that of $d_{xz}(d_{yz})$ states (~7.6 eV) gets larger than that for cubic SRO (~7.1 eV), as consequences of the change of the bond lengths between Ru ions with in-plane and out-of-plane O ions.

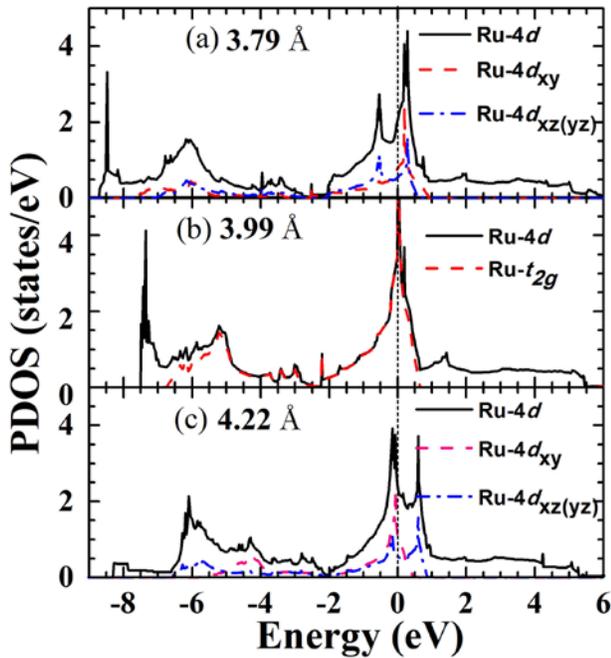

FIG.4. (color online) Calculated non-spin-polarized density of states of Ru 4$d$ states in SRO with in-plane lattice parameter: (a) 3.79 Å, (b) 3.99 Å, (c) 4.22 Å. Red dashed and blue dash dotted lines in (a) and (c) are Ru $d_{xy}$ and $d_{xz(yz)}$ states, respectively. Red dashed lines in (b) are Ru $t_{2g}$ states. The black short dashed line marks the Fermi level.

To explain the energy splitting between the spin-up and spin-down states and analyze the magnetic transition under strain effect, we then adopt the Stoner model[36]. The condition of the stability of the itinerant ferromagnetic state is:

$$I \cdot N_F > 1, \qquad (1)$$

where $I$ is the Stoner parameter and $N_F$ is the nonmagnetic density of states at the Fermi level. From Eq. (1), it can be inferred that a large value of $N_F$ or $I$ favors the ferromagnetic state. PDOS plots of Ru 4$d$ states under different in-plane lattice constants have been obtained using non-spin-polarized DFT calculations, which are shown in Fig. 4. It is clear that when the lattice constant $a$ is 3.99 Å, a very sharp PDOS peak is located exactly at the Fermi level. As a consequence, the cubic SRO has the strong ferromagnetism. When $a$ is 3.79 Å or 4.22 Å, the peak moves away from the Fermi level. Specifically, the peak of the PDOS moves above (below) the Fermi level under compressive (tensile) strain. For $a$=3.79 Å, the deviation is more pronounced, as both $d_{xy}$ and $d_{xz(yz)}$ states experience band-width expansion, which naturally decrease the PDOS at Fermi level ($N_F$) and the Stoner parameter $I$. Due to this effect, all three $d$ states tend to be non-magnetic, resulting in zero magnetic moment of the system. Whereas for $a$=4.22 Å, the band-widths of $d_{xz(yz)}$ states increase as a result of the enhanced Ru-O hybridization along the $z$ direction, and due to the same reason their spin-up and spin-down channels are almost equally occupied. The $d_{xy}$ states, however, have smaller band-width than that of cubic SRO, so it still has PDOS peak very close to the Fermi level, which favors a ferromagnetic state. Therefore, for $a$=4.22 Å, SRO keeps the weak ferromagnetism, and the major contribution to the magnetic moment comes from the $d_{xy}$ state. All these demonstrate that the Stoner model is capable of explaining the magnetic transitions of SRO under epitaxial strain.

In the above calculations, we assume the structures of SRO possess inversion symmetry, i.e., the Ru atom locates at the center of the unit cell and no rumpling between the in-plane anions and cations. To further study the influence of the epitaxial strain on the structural instabilities of SRO film, we then carry out the phonon frequencies calculations of the tetragonal SRO using the first-principles frozen-phonon method. The results show that when $a$ < 3.74 Å or $a$ > 4.10 Å, SRO has imaginary phonon frequencies at Γ point. To be specific, same as the case of perovskite ferroelectric oxides, tetragonal SRO experiences a splitting of the cubic TO mode into a single $A_{2u}$ mode polarized perpendicular to the $xy$ plane, and a twofold degenerate $E_u$ mode polarized in the $xy$ plane. With the increase of compressive strain, the $A_{2u}$ mode gradually becomes *soft*, i.e, its frequency decrease. After about 3.74 Å (~ -6% strain), the frequency becomes imaginary, indicating the occurrence of the unstable vibration mode. Whereas with the increase of tensile strain, the $E_u$ mode becomes the *soft* mode. At about 4.10 Å, the phonon instability at Γ point appears, suggesting the ground state of SRO no longer has inversion center.

To search for the stable state of SRO with phonon instability at Γ point, we have moved the atomic positions of Ru and O atoms according to the unstable phonon mode (Sr atom is fixed as reference). In Fig. 5, the so obtained curves of energy evolution with the softmode amplitude, here characterized as Ru offcenter displacement, clearly demonstrate double-well profile, a general feature of ferroelectric materials. Note that potential well-depth is also of the same magnitude (~10 meV) of the typical ferroelectric oxides. Especially in tensile strain, e.g., $a$=4.22 Å, the Ru off-center displacement is as large as 0.06 Å, a value can be unambiguously observed experimentally. Indeed, the movement of O atoms is even larger.

What more interesting is, as for $a$=4.22, the magnetic moment of SRO will change with the softmode amplitude, as shown in the inset of Fig. 5(b). It changes from ~ 0.7 $\mu_B$ at the center symmetric position to ~ 0.3 $\mu_B$ at the *polarized* position. Detailed analysis reveals that this is due to the increased hybridization of Ru with planar O atoms, similar as the situation of strain effect discussed earlier.

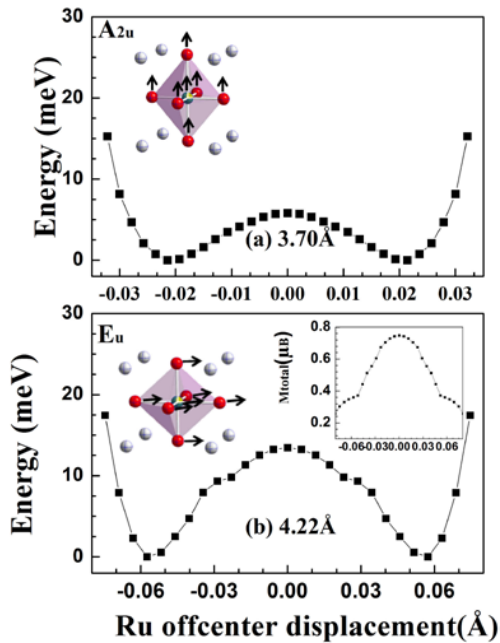

FIG.5. (color online) Total energy as a function of Ru offcenter displacement for lattice distortion according to the unstable phonon mode of SRO: (a) $A_{2u}$ mode for the compressive strain (a=3.70 Å), (b) $E_u$ mode for the tensile strain (a=4.20 Å). The inset of (b) shows the magnetic moment change with the softmode amplitude.

Finally, we have also considered the electron-electron correlation effects on the above findings, even though SRO should be relatively weak correlated system due to large spatial extent of the 4$d$ orbitals in the ruthenates.[8] We adopt GGA+$U$ method with the effective on-site Coulomb interaction energy for Ru 4$d$ electrons $U_{eff}$=0.6 eV to treat the correlation effect.[8] It is found that the consideration of correlation effect does not affect the major findings of our above studies, just the onset values of the epitaxial strains for the occurrence of magnetic transitions and phonon instabilities shift a little. For example, the unstable $E_u$ mode mode will occur at in-plane lattice parameter $a$~4.12 Å, compared with 4.10 Å in the case of no correlation effect taken into account.

In conclusion, we have studied the magnetic and dynamical properties of tetragonal SRO under the influence of epitaxial strain by using DFT calculations. Three magnetic states, i.e., non-magnetic, strong ferromagnetic, and weak ferromagnetic are found with the application of different epitaxial strain. We find that the Stoner model could provide a reasonable explanation on such phenomena. In addition, calculations also reveal that the metallic SRO may exhibit *ferroelectric-like* soft mode behavior under certain strain. This could have influence on the properties of the oxide superlattices consisting of SRO films.

This work was supported by the 973 Program Nos. 2013CB922300, 2011CB922101 and 2010CB923404, the NSF of China (Grant No. 61125403, 50832003, 11004211, 91122035, 11174124), PCSIRT, NCET, Fundamental Research Funds for the central universities (ECNU), and the ECNU Fostering Project for Top Doctoral Dissertations. Computations were performed at the ECNU computing center.
*Electronic address: sjgong@ee.ecnu.edu.cn